\documentclass[useAMS,usenatbib]{mn2e}

\usepackage{graphicx}
\usepackage{natbib}
\usepackage{aas_macros}

\title[The shortest period contact binary]
{The shortest period field contact binary\thanks{Based on the data 
obtained at the David Dunlap Observatory, University of Toronto
and the All Sky Automated Survey.}}

\author[Slavek M. Rucinski and Theodor Pribulla]
{Slavek M. Rucinski\thanks{E-mail: rucinski@astro.utoronto.ca} 
and Theodor Pribulla\thanks{On leave from the 
Astronomical Institute of the Slovak Academy of Sciences, 
059\,60 Tatransk\'{a}, Lomnica, The Slovak Republic; 
E-mail: pribulla@ta3.sk} \\
Department of Astronomy, University of Toronto, 
50 St.~George St., Toronto, Ontario, Canada M5S~3H4
}

\date{Accepted --.
      Received -- ;
      in original form --}

\pubyear{2008}

\begin{document}

\maketitle

\label{firstpage}

\begin{abstract}
Photometric and spectroscopic results for the
contact binary GSC~01387--00475 (ASAS~083128+1953.1)
are presented. The existence of this binary with 
the orbital period of $P=0.2178$ d strengthens 
the argument that the cut-off
of the period distribution for contact binaries -–
until now defined by CC~Comae -- is very sharp. 
The only case of a still shorter period is known 
in a globular cluster where more compact 
contact configurations are in fact expected. 
While the spectroscopic orbit of GSC~01387--00475 
is well defined, the low orbital inclination of
the binary and the presence of a spectroscopic 
companion contributing about 1/3 of the total light 
conspire to reduce the photometric variability to
$\simeq 0.09$ mag. The photometric data are 
currently inadequate to identify the source of the 
small amplitude 
(0.02 -- 0.03 mag) intrinsic variability of the system. 
\end{abstract}


\section{Introduction}  
\label{intro}

The period distribution of contact binaries is known to have 
a sharp cut-off at short periods \citep{Rci1992a,Rci2007}.
Currently, the shortest-period, 
well researched, field system is CC~Comae with $P=0.2207$ d; 
for the most recent spectroscopic data and references on
CC~Com, see \citet{ddo12}. 
The period distribution cut-off does not have
a fully satisfactory explanation. A recent suggestion 
\citep{Step2006} sees it as a result of a very strong 
decrease in the efficiency of the angular-momentum ($H$) loss 
with the decreasing mass ($M$) along the Main Sequence
for the ``saturated'' magnetic-activity
case. Specifically, $-dH/dt \propto \omega R^2 M \propto M^3$, 
since for the lower MS the radius scales with the mass, 
$R \propto M$. The cut-off would be
then a result of the finite age of the binary population
forming contact systems. 

The All Sky Automated Survey  
ASAS \citep{Poj1997,Poj2004,Poj2005,BP2006} has 
been contributing greatly to discovery of many contact binaries
in the still not fully explored magnitude range 8 -- 13 mag,
particularly in the Southern hemisphere where hundreds of
contact binaries remain to be detected. 
Out of seven variable stars discovered by the ASAS and originally
classified as contact binaries with periods shorter than the
CC~Com period \citep{Rci2006}, six
have been subsequently shown \citep{Rci2007} not to be
binaries at all. In this paper we describe the only
system which remained, ASAS~083128+1953.1. With the orbital
period of 0.217811 d, which is 
by 1.3\% shorter than that of CC~Com ($P=0.220686$ d), 
this is the shortest-period sky-field 
contact binary currently known (an even tighter system
exists in a globular cluster, see Section~\ref{concl}).

Instead of the ASAS designation, we 
use the Guide Star Catalogue (also Tycho) name of the star,
GSC~01387--00475. 
In addition to our spectroscopic observations obtained at 
the David Dunlap Observatory (DDO), we present an
analysis of photometric $V$ and $I$ 
data from the ASAS survey, supplemented by white-light 
(un-filtered) photometric observations from DDO. 

Contact binary stars have been a subject of several recent
spectroscopic studies at the DDO,
leading to over one hundred radial-velocity
orbits (see the last published paper DDO-12, \citet{ddo12}).
In the DDO series, we normally group binaries into batches of ten;
this is done because of the similarity of many targets as well as
of our desire not to further inflate the  
close-binary star literature. 
This paper is an exception as it describes only one 
object. However, we feel that the star deserves a speedy and
separate publication because of its special position at the
very end of the orbital period sequence.

\section{Literature information}
\label{literat}

GSC~01387--00475 is located at 
J(2000) 08:31:27.88, +19:53:03.5.
Its Tycho-2 \citep{Tycho2}
mean magnitudes are: $B_T = 11.922$ and $V_T = 10.712$
resulting in $B-V = 1.03$ and the spectral type on the
Main Sequence of K4V. This is inconsistent 
with $V-I = 1.25$ from the ASAS 
project (Section~\ref{phot}) which suggests $B-V = 1.12$ 
and thus K5V. Our own direct spectral classification
from the blue classification spectra (a window of 
about 650 \AA\ wide centred on 4200 \AA) 
is a well defined K3V while the
2MASS magnitudes result in $J-K = 0.72$ corresponding
to the spectral type K5V. The spread in the
estimated spectral types may result from the presence
of the newly detected
spectroscopic companion (Section~\ref{comp}).

The trigonometric parallax of GSC~01387--00475 is unknown.
The extrapolated to short periods $M_V$ calibrations 
\citep{Rci2000} give substantially
different values of $M_V = +6.17$ and $+7.16$, for the
assumed $B-V=1.03$ and $V-I=1.25$, respectively; this is
based on the assumed $V_{max}=10.33$ and the binary contribution
of 2/3 to the total brightness of the system.  
The extrapolated $M_V$ calibrations are very approximate and 
this very system may eventually help in establishing them
for the shortest periods. Because of the uncertain colour
indices and the calibrations, the implied distance estimates 
span a range between 53 pc and 84 pc.  

The proper motion of GSC~01387--00475 is 
$-11.1 \pm 1.6$ mas/yr and $+5.5 \pm 1.6$ mas/yr
in the two tangential sky directions \citep{Tycho2}.
Assuming the convention of
\citet{JS1987}, the solar motion $UVW_\odot = +9, +12, +7$,
and the radial velocity of the binary 
$V_0=+22.21$ km~s$^{-1}$ (Section~\ref{spec}),
the space velocity components are $UVW = -10.3, +5.6, +16.5$
km~s$^{-1}$ for the distance of 53 pc and 
$UVW = -11.4, +6.5, +15.4$
km~s$^{-1}$ for the distance of 84 pc. Therefore, the 
star has a rather moderate space velocity. 

A marginally significant X-ray source 1RXS~J083127.8+195301
(ROSAT flux: $0.102 \pm 0.020$ cts/s, \citet{Voges1999})
is located within 6 arcsec of the optical location indicating
a positional coincidence. 
GSC~01387--00475 has not been recognized before as a visual binary,
there are no obvious visual companions close
to it and none was noticed in the 
slit view during the spectroscopic observations.

\begin{figure}
\begin{center}
\includegraphics[width=70mm]{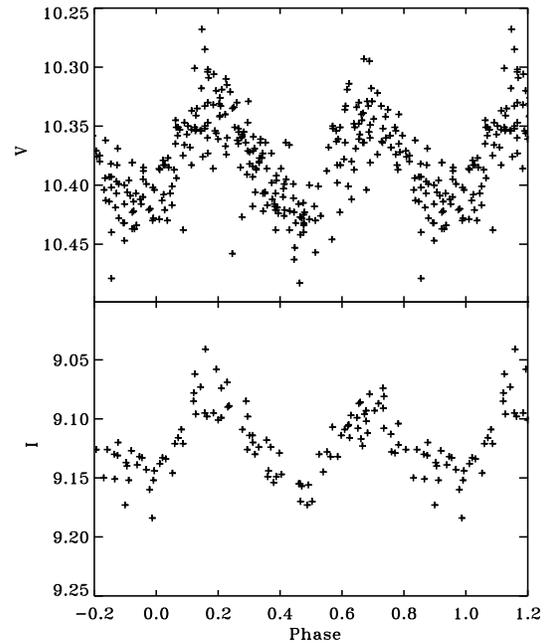}
\caption{The $V$ and $I$ photometric band ASAS data
for GSC~01387--00475 versus the orbital phase
calculated using our spectroscopic conjunction (eclipse)
prediction as in Section~\ref{spec}.}
\label{ASASph}
\end{center}
\end{figure}

\begin{table}
\begin{small}
\caption{ASAS $V$ and $I$ photometry of GSC~01387--00475.
\label{tab_asas}}
\begin{center}
\begin{tabular}{ccc}
\hline
HJD & $V$ or $I$ & Filter\\
\hline
2452622.7490 & 10.335 & $V$ \\
2452624.7771 & 10.411 & $V$ \\
2452635.7268 & 10.374 & $V$ \\
2452635.7268 & 10.322 & $V$ \\
2452637.7681 & 10.375 & $V$ \\
\hline
\end{tabular}
\end{center}
\end{small}
\end{table}

\begin{figure}
\begin{center}
\includegraphics[width=80mm]{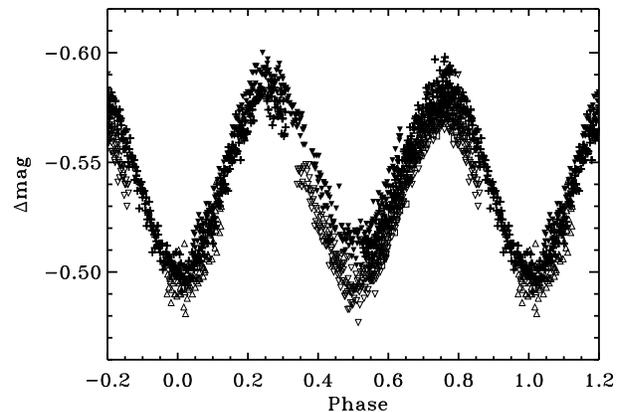}
\caption{The unfiltered DDO photometry of 
GSC~01387--00475 versus the orbital phase
calculated using the time of the spectroscopic conjunction 
(Section~\ref{spec}). Different symbols are used for
different nights.
}
\label{DDOph}
\end{center}
\end{figure}

\begin{table}
\begin{small}
\caption{DDO differential, unfiltered photometry of GSC~01387--00475
relative to GSC~01387--00510.
\label{tab_ddo}}
\begin{center}
\begin{tabular}{cc}
\hline
HJD & $\Delta m$ \\
\hline
2454422.8607 & $-0.584$ \\
2454422.8610 & $-0.589$ \\
2454422.8614 & $-0.587$ \\
2454422.8617 & $-0.590$ \\
2454422.8621 & $-0.596$ \\
\hline
\end{tabular}
\end{center}
\end{small}
\end{table}

\section{Photometry}  
\label{phot}

The ASAS data 
\citep{Poj2004,Poj2005,BP2006}\footnote{For details, see:
http://www.astrouw.edu.pl/$\sim$gp/asas/asas.html and 
http://archive.princeton.edu/$\sim$asas/ }, 
for GSC~01387--00475 
have been collected during 6 consecutive seasons 
(2003 -- 2008) in the $V$ band and during 3 seasons (2003,
2005, 2006) in the $I$ band. 
Only the best, grade ``A'', observations were used.
The data are listed in the
on-line Table~\ref{tab_asas} and consist of 290 
observations in the $V$ filter and 102 observations
in the $I$ filter. 

When plotted with the 
spectroscopic time elements ($T_0 = 2452623.1423$,
$P= 0.217811$ d; Section~\ref{spec}), the phased observations 
(Figure~\ref{ASASph}) show a light curve of a small
amplitude of about 0.09 mag, but with a 
relatively large scatter, particularly in $V$, of
about 0.02 -- 0.03 mag. The scatter is larger  
than expected for the ASAS data at the same 
magnitude level \citep{Poj2004} where it should be typically
below $0.015 - 0.02$ mag.  
The scatter is not due to any period changes as the eclipse
times do not show any systematic shifts
during the six year span of the ASAS observations.

Because of the northern sky
location observed from the southern observatory
of Las Campanas, the ASAS data points were acquired 
relatively infrequently, with typical spacing of  
2 -- 3 days, with some gaps lasting as long as 
several days. The large scatter can therefore
come from some other source of variability, either
due to the binary itself or to its spectroscopic companion
(Sections~\ref{spec} -- \ref{comp}). 

While conducting the spectroscopic observations at
the DDO, simultaneous unfiltered photometry of
GSC~01387--00475 
was obtained on 6 nights (18 Nov.\ 2007 to 25 March 2008) 
using the 15 cm finder of the main 1.88m telescope. The
photometric sampling was more rapid, when compared 
with that of the ASAS data, with the typical spacing
between observations of one minute. The observations
(the on-line Table~\ref{tab_ddo})
clearly show a well defined, low amplitude
($\Delta m = 0.09$) binary light curve 
on individual nights. The small differences between the
nights of typically 0.01 -- 0.02 mag (Figure~\ref{DDOph})
do not have an explanation and may be 
relatively slow, although their time scale cannot be
established from our very infrequent nightly
observations\footnote{The long duration of the DDO
observations was not planned and was 
entirely due to the poor weather during the
2007/2008 winter. 
At DDO, we normally try to obtain a reasonable 
spectroscopic phase coverage in as 
short a time span as possible.}. 
Unfortunately, we cannot exclude a possibility
of some instrumental causes because the 15 cm finder camera 
was never meant to serve as a precise photometric 
instrument and its long-term stability is
entirely unknown. Therefore, we are not sure if the
photometric scatter in the ASAS data can be explained
by the night-to-night shifts observed during the DDO
observations. 

The DDO observations shown in Figure~\ref{DDOph}
are expressed as magnitude differences relative to the
comparison star GSC~01387--00510. Three eclipses
were observed sufficiently to establish times of
minima at HJD 2454422.9169(5), 2454525.6136(3), and
2454550.6627(3); the standard errors are expressed  
in units of the last decimal place and are given in
parentheses. Based on these, the value of the
photometric epoch is by 0.0018 d later than based
on the spectroscopic results (Table~\ref{tab_param}). 

While the small amplitude of the binary variations 
is partly due to the presence of the third
star in the system which ``dilutes'' the binary star
signal (Section~\ref{comp}), most of the amplitude
reduction must come from the low orbital inclination. 
Assuming the contact model with the 
spectroscopic mass ratio of
$q = 0.47$ (Section~\ref{spec}) and the 2/3 light 
contribution of the binary 
to the total brightness (Section~\ref{comp}),
we estimate the orbital inclination at $i = 42 \pm 3$ 
degrees; the uncertainty includes an unknown 
(and presumed small) degree-of-contact of the binary.

\section{Spectroscopy}
\label{spec}

Spectroscopic 
observations of GSC~01387--00475 were obtained using the slit 
spectrograph in the Cassegrain focus of 1.88m 
telescope of the David Dunlap Observatory. The spectra were 
taken in the window of 240 \AA\ around the Mg~I 
triplet (5167, 5173 and 5184~\AA) with an effective 
resolving power of about $12,000$ provided by a diffraction 
grating with 2160 lines/mm and 240~$\mu$m slit at the scale of 
0.117 \AA/pixel. The efficient program of
\citet{pych2004} was used for removal of cosmic rays from the
2-D spectral images. One-dimensional spectra were extracted 
by the usual procedures within the IRAF 
environment\footnote{IRAF is distributed by the National 
Optical Astronomy Observatories, which are operated by 
the Association of Universities for Research in Astronomy, 
Inc., under cooperative agreement with the NSF.} 
after the bias subtraction and the flat field division.
The exposure times were always 500 seconds long, which
corresponds to 2.6\% of the orbital period.

\begin{figure}
\begin{center}
\includegraphics[width=65mm]{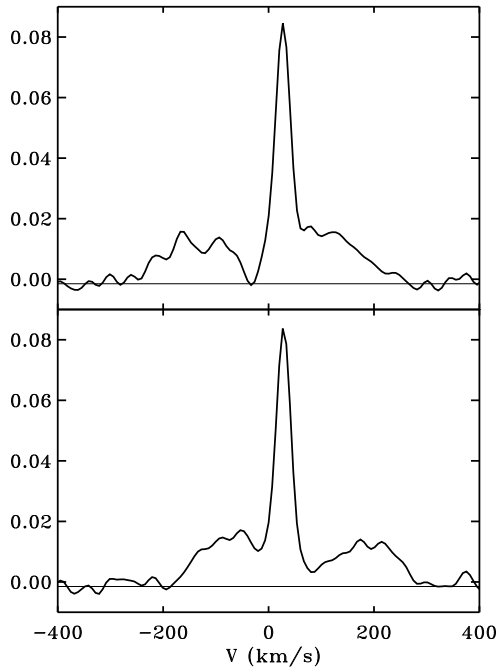}
\caption{The broadening functions for two orbital phases
of GSC~01387--00475: the upper panel for 0.754 and the lower
panel for 0.257. The vertical scale is in basically arbitrary units
of the BF intensity per 6.7 km~s$^{-1}$ pixel; for a
perfect spectral match of the template, an integral over 
the BF in such units would give exactly unity.}
\label{BFs}
\end{center}
\end{figure}

Broadening functions (BFs), which can be considered as
images of the binary system in the radial velocity domain
(Figure~\ref{BFs}) were extracted by the method described in 
\citet{Rci1992b} and \citet{Rci2002} (see also other papers
of the DDO series, e.g., \citet{ddo12}). As templates
served stars HD~3765, K2V, $V_r = -62.3$ km~s$^{-1}$ and
HD~65583, G8V, $V_r = +13.2$ km~s$^{-1}$. The integrals of the
BFs are substantially larger than unity, of 1.09 and 1.46,
respectively, implying an increased combined strength of the lines, 
in agreement with the estimated spectral type of the binary 
of K3/5V.  

The phase system that we used assumes the zero phase at the time
of the eclipse of the more massive component. It is not clear
if this component has a larger surface brightness; 
the light curves are not defined well enough to distinguish
which of the eclipses is the deeper one. The photometric scatter
appears to be larger during the secondary minima (Figure~\ref{DDOph}),
but this may be due to the unevenly dense photometric 
phase coverage.

\begin{figure}
\begin{center}
\includegraphics[width=80mm]{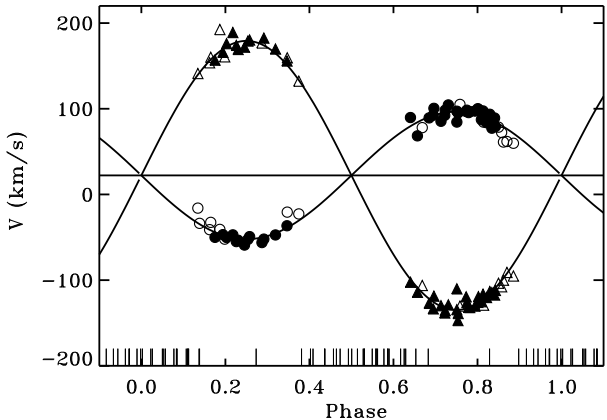}
\caption{The radial velocity orbit for the binary star
GSC~01387--00475. The phases are counted from the conjunction
with the more massive component behind. Observations
carrying a reduced weight of 1/2 are marked by open
symbols. 
The orbital elements are given in Table~\ref{tab_param}.
Phases of unused spectra with blended or poorly 
defined signatures of both components in the BFs are 
marked at the bottom margin of the figure.}
\label{RVbin}
\end{center}
\end{figure}

\begin{table}
\begin{small}
\caption{The radial velocities of all three components of
GSC~01387--00475. 
Weights $w_i$ refer to the quality of radial velocity 
determinations.
\label{tab_rv}}
\begin{center}
\begin{tabular}{cccccc}
\hline
HJD & $V_1$ & $V_2$ & $V_3$ & $w_1$ & $w_2$ \\
\hline
2454422.8610  &   95.25 &  $-139.00$ &  26.68 & 1.0 & 1.0\\
2454422.8670  &   96.72 &  $-131.92$ &  25.89 & 1.0 & 1.0\\
2454422.8730  &   87.27 &  $-125.25$ &  25.30 & 1.0 & 1.0\\
2454422.8803  &   79.16 &  $-111.99$ &  26.39 & 1.0 & 1.0\\
2454422.8863  &   61.79 &  $ -91.06$ &  27.60 & 0.5 & 0.5\\
\hline
\end{tabular}
\end{center}
\end{small}
\end{table}

The broadening functions are well defined 
with a prominent signature of a third, slowly rotating component. 
The radial velocities of all three components 
(the on-line Table~\ref{tab_rv}) were measured 
using our standard approach developed in the DDO series
\citet{ddo12}: We
fitted the BFs by three Gaussians, subtracted the Gaussian of 
the third component, and then fitted a double rotational
profile to the residual binary peaks. 

The radial-velocity orbit of GSC~01387--00475 is
shown in Figure~\ref{RVbin}. Given the short exposures and 
the relative faintness of the system, 
the orbit is surprisingly well
defined. The spectroscopic parameters 
are listed in Table~\ref{tab_param}. With the orbital
inclination of $i = 42 \pm 3$ degrees (Section~\ref{phot}),
the total mass of the binary, 
$M_1+M_2=0.94 \pm 0.16$ M$_\odot$, appears to be 
reasonable for a contact binary of the K3/5V spectral type;
the light curves are currently too poor to lift 
the large inclination uncertainty in the sum of the
masses. The estimated total mass appears to be
very similar to that of CC~Com,  
$M_1 + M_2 = 1.083 \pm 0.012\,M_\odot$; the 
the mass ratio, $q = 0.474 \pm 0.008$ is also similar,
CC~Com: $q=0.527 \pm 0.006$ \citep{ddo12}.  

\begin{table}
\begin{small}
\caption{Parameters of the spectroscopic orbit of 
GSC~01387--00475.
\label{tab_param}}
\begin{center}
\begin{tabular}{cccc}
\hline
Parameter & Unit & Value & Std. Err. \\
\hline
$P$     & day         & 0.217811 & fixed \\
$T_0$   &             & 2452623.1423 & 0.0005 \\
$V_0$   & km~s$^{-1}$ & 22.21 & 0.68 \\
$K_1$   & km~s$^{-1}$ & 74.40 & 0.89 \\
$K_2$   & km~s$^{-1}$ & 157.00 & 1.25 \\
$(M_1+M_2) \sin^3 i$ & $M_\odot$ & 0.280 & 0.008 \\
$q = M_2/M_1$     &             & 0.474 & 0.008 \\
\hline
\end{tabular}
\end{center}
\end{small}
\end{table}

\section{The spectroscopic companion}
\label{comp}

The third component of GSC~01387--00475 was analyzed as
a single star from the BFs, after removal of the close-binary
signature. The radial velocities of the companion were
measured only for the best quality BFs.  
The error per observation was larger 
than typically for single,
sharp-line stars ($<1.0$ km~s$^{-1}$) and equalled 1.27 km~s$^{-1}$
which may be partly explained by the variable ``cross-talk''
from the underlying, rapidly changing 
contribution of the binary in the BFs
(see Figure~\ref{BFs}). The average radial velocity,
$V_3 = 27.42 \pm 0.18$ km~s$^{-1}$ (the error
evaluated assuming the constant value),   
is similar to, but significantly different from the centre 
of mass velocity of the binary of $V_0=+22.21 \pm 0.68$ km~s$^{-1}$.
This measurable difference may suggest a moderately short orbital
period of the third component relative to the common mass
centre. 

The value of the rotation broadening is indeterminable from the
broadening functions. We can give only an upper limit of
$V_3 \sin i < 15$ km~s$^{-1}$.

The luminosity ratio of the third star relative 
to the binary, estimated from the integrated 
peaks in the broadening functions, such as in Figure~\ref{BFs}, 
is $L_3/L_{12} = 0.546 \pm 0.015$. This value is most
likely an over-estimate because of the biased 
continuum normalisation during rectification 
of the spectra: the normalisation applies to the real 
continuum for the sharp-line companion and to the 
pseudo-continuum for the heavily blended spectra of the 
close binary. An estimate including this systematic effect is
$L_3/L_{12} = 0.50 \pm 0.05$. Thus, the three
components of GSC~01387--00475 appear similar in their
luminosity so that the third component is also a mid-K type
dwarf.

\section{Conclusions}
\label{concl}

The existence of GSC~01387--00475  confirms that
the short period cut-off of the contact 
binary period distribution at 0.22 days 
is sharp and well defined. 
CC~Com is definitely not an exception.
The existence of a still shorter period contact binary
V34 identified by \citet{Waldr2004} in the globular 
cluster 47~Tuc with the period of $P=0.2155$ d is
consistent with both current
explanations, the expected smaller dimensions of
contact binaries in globular clusters \citep{Rci2000} and with
the more advanced age of such binaries compared with the
galactic field objects \citep{Step2006}.

GSC~01387--00475 may be slightly less massive than CC~Com and thus
slightly more extreme in its properties.
However, as an eclipsing and spectroscopic binary,
GSC~01387--00475 is a less convenient object for detailed studies than 
CC~Comae because it has a lower orbital inclination angle
and hence does not show total eclipses as does CC~Com. 
Additionally, its light curve may be subject to unexplained
intrinsic variations, possibly with a long time scale of
weeks or months. The variability may be due to the binary
itself or to its spectroscopic
companion which contributes about 1/3 of the total light
of the system. The existence of the companion one more
time confirms the recent results that practically all contact 
binaries have companions on wide orbits \citep{3rd-1,3rd-2,3rd-3}.

\medskip

The authors would like to express their thanks 
to Dr.\ G.\ Pojma\'{n}ski for making ASAS such an 
useful variable-star study tool and to Mr.\ B.\
Pilecki for the extraction of the photometric data from the
ASAS database and the attempts to find long period trends
in these data. Dr. K. St\c{e}pie\'{n} kindly read the paper
and asked pointed questions.

Research support from the Natural Sciences and Engineering 
Council of Canada is acknowledged with gratitude. 
The research made use of the SIMBAD database, operated at the CDS,
Strasbourg, France and accessible through the Canadian
Astronomy Data Centre, which is operated by the Herzberg Institute of
Astrophysics, National Research Council of Canada.

\end{document}